\documentclass[english,prl]{revtex4}
\usepackage[T1]{fontenc}
\usepackage[latin9]{inputenc}
\usepackage{bm}
\usepackage{amsmath}
\usepackage{amssymb}
\usepackage{wasysym}
\usepackage{graphicx}
\usepackage{esint}
\usepackage{bm}
\newcommand{\beq}{\begin{equation}}
\newcommand{\eeq}{\end{equation}}
\newcommand{\beqa}{\begin{eqnarray}}
\newcommand{\eeqa}{\end{eqnarray}}
\makeatletter

\@ifundefined{textcolor}{}
{%
 \definecolor{BLACK}{gray}{0}
 \definecolor{WHITE}{gray}{1}
 \definecolor{RED}{rgb}{1,0,0}
 \definecolor{GREEN}{rgb}{0,1,0}
 \definecolor{BLUE}{rgb}{0,0,1}
 \definecolor{CYAN}{cmyk}{1,0,0,0}
 \definecolor{MAGENTA}{cmyk}{0,1,0,0}
 \definecolor{YELLOW}{cmyk}{0,0,1,0}
}

\makeatother
 
\usepackage{babel}
\begin{document}
\bibliographystyle{naturemag}

\title{Raise and collapse of strain-induced pseudo-Landau levels in graphene}

\author{Eduardo V. Castro,$^{1,2}$ Miguel A. Cazalilla,$^{3,4}$ Mar\'ia A. H. Vozmediano $^5$}
\affiliation{$^1$ CeFEMA, Instituto Superior T\'{e}cnico, Universidade de 
Lisboa, Av. Rovisco Pais, 1049-001 Lisboa, Portugal}
\affiliation{$^2$Beijing Computational Science Research Center, Beijing 100084, China }
\affiliation{$^3$ Department of Physics, National Tsing Hua University, Hsinchu City, Taiwan}
\affiliation{$^4$ National Center for Theoretical Sciences (NCTS), Hsinchu City, Taiwan}
\affiliation{$^5$ Instituto de Ciencia de Materiales de Madrid, and CSIC, Cantoblanco, 28049 Madrid, Spain}

\begin{abstract}
{\bf Lattice deformations couple to the low energy electronic excitations of graphene as vector fields similar to the electromagnetic potential \cite{SA02b,VKG10}.  The suggestion that certain strain configurations would be able to induce pseudo landau levels in the spectrum of graphene \cite{GKG10,GGKN10}, and the subsequent experimental observation of these \cite{LBetal10} has been one of the most exciting events in an already fascinating field. It opened a new field of research ``straintronics" linked to new applications, and had a strong influence on  the physics of the new Dirac materials in two and three dimensions \cite{Amorim16}. The experimental observation of pseudo landau levels with scanning tunnel microscopy presents nevertheless some ambiguities. Similar strain patterns  show different images sometimes difficult to interpret. 
In this work we strain the analogy of the pseudo versus real electromagnetic fields, as well as the fact that graphene has a relativistic behavior and show that, for some strain configurations, the deformation potential acts like a parallel electric field able to destabilize the Landau level structure via a mechanism identical to that described in \cite{LSB07,PC07} for real electromagnetic fields. The underlying physics is the combination of Maxwell electrodynamics and special relativity implying that different reference frames observe different electromagnetic fields \cite{LSB07}. The mechanism applies equally if the electric field has an external origin, which opens the door to an electric control of the giant pseudomagnetic fields in graphene.}

\end{abstract}
\maketitle

As it is well known, lattice deformations couple to the low energy electronic excitations of graphene as vector fields with a minimal coupling similar to that of the electromagnetic potential. This old observation done first in nanotubes  \cite{KM97,SA02b}, acquired an unexpected development in graphene when it was suggested that certain strain configurations would induce a constant pseudomagnetic field in some regions of the sample able to break the spectrum into pseudo landau levels (PLL) \cite{GKG10,GGKN10}. The subsequent experimental observation with a scanning tunneling microscopy (STM) experiment  of LL corresponding to pseudomagnetic fields of the order of 300~T \cite{LBetal10} has been one of hallmarks in the physics of graphene. The possibility to manipulate the electronic properties with lattice deformations is on the basis of many suggested technological applications and opened the new field of graphene straintronics \cite{KJetal12,SSL16}.

Although PLL have been established in many different experiments in graphene \cite{MH05,YS12,LBetal15} and also in optical or artificial lattices \cite{JMetal14,GMW2012,PG13}, the STM images are often difficult to interpret. It is observed that the spectral lines for very similar deformation patterns  differ notably from each other and sometimes PLL are not observed in configurations that would correspond to high enough pseudomagnetic fields. 
We will address this problem on the basis of another beautiful property of graphene under strain that strengthens the relativistic nature of the system.

The other peculiar property of the graphene electronics is the ultra-relativistic nature of its charge carriers despite the fact that their Fermi velocity $v_F$ is of the same order of magnitude as that of other metals. It is the masslessness what implies relativistic dynamics yet defying the relativistic principle that massless particles move at the speed of light. As learned in elementary courses of special relativity, that electromagnetic fields does not look the same in different reference frames. In particular, a Lorentz transformation with velocity $v$ perpendicular to a field configuration $E\perp B$ can rotate away the electric field leaving only a magnetic field if $v=E/B$. This fact was advocated in Ref.~\cite{LSB07} to show that the  Landau levels of graphene would collapse in the presence of an in-plane electric field above a critical value $E_c=v_F B$. The same result was obtained in Ref.~\cite{PC07} using an alternative approach. Since the Fermi velocity in graphene takes the value $v_F \simeq 10^6~\mathrm{m/s}$, this gives a critical field $E_c \approx 10^{-3}B\mathrm{[T]~V/nm}$ for a magnetic field $B$ in Tesla. In conventional quantum Hall experiments this field is absent, even though the corresponding critical field is well within reach for typical magnetic fields. The situation is very different in the case of strain induced pseudomagnetic fields as we will see in what follows. We will show that, for certain strain patterns giving rise to a uniform pseudomagnetic field, an unavoidable parallel electric field induced by deformation potential can destabilize the Landau level structure via a mechanism identical to that described in Refs.~\cite{LSB07,PC07}.

\section*{Graphene in perpendicular electric and magnetic fields.}
\begin{figure}
\includegraphics[width=10cm]{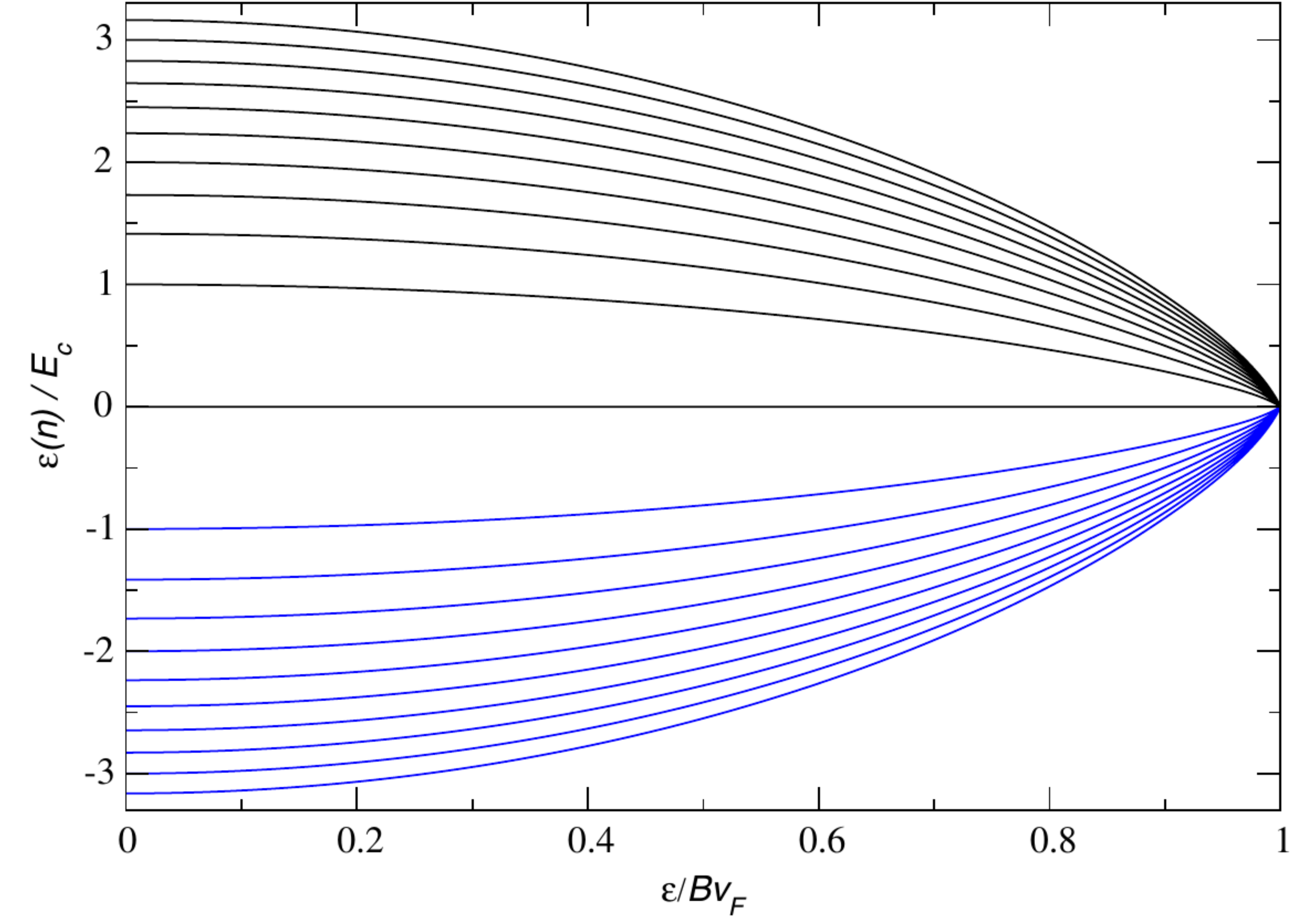}
\caption{Evolution of the LL with the ratio $E/v_F B$. $E_c=\hbar \omega_c$ is the cyclotron energy. Fig. taken from ref. \cite{PC07}.}
\label{fig:E1}
\end{figure}  
The Landau level spectrum of graphene in a perpendicular magnetic field is 
\beq
\epsilon(n)=\pm\hbar \omega_c\sqrt{n},\hspace{0.5cm} n=0,1,2,\dots
\eeq
where the cyclotron frequency is $\omega_c=\sqrt{2}v_F/l_B$, with $l_B = \sqrt{\hbar /(eB)}$ the magnetic length~\cite{Goerbig11}. 
In the presence of an in-plane electric  field $E$ the spectrum is modified to
\beq
\epsilon(n)=-eEkl_B^2\pm\hbar\Omega_c\sqrt{n},
\eeq
where, apart from the conventional energy shift $eEkl_{B}²$ due to the position of the Landau state ($k$ is momentum along the longitudinal direction), there is a new cyclotron frequency $\Omega_c$ given by \cite{PC07},
\beq
\Omega_c=\sqrt{2}\frac{v_f}{l_B}[1-E^2/(B^2v_F^2)]^{3/4}.
\eeq
When the electric field $ E $ approaches $ v_f B$ from below, the LL become closer and collapse at the critical value $E = v_f B$ (Fig. \ref{fig:E1}).
This effect was described in a very elegant manner in ref. \cite{LSB07} as a consequence of the relativistic nature of the graphene system and the mixing
of electric and magnetic fields in moving frames of reference. The LL collapse has been observed  in~\cite{kimLevitovLLCollapse2011} for graphene under a top gate.

Even before the critical field is reached and LL collapse observed, the Hall conductivity quantization breaks down for electric fields above $E=\sqrt{2} v_F B l_B/L$, where $L>l_B$ is the width of the Hall bar. This happens because above this field extended bulk LL always cross the Fermi level, which for graphene occurs when $E\gtrsim 0.036 \sqrt{B\mathrm{[T]}} /L\mathrm{[nm]}~\mathrm{V/nm}$, with magnetic field $B$ in Tesla and length $L$ in nanometers.  Nevertheless, local spectroscopic probes like STM still provide a spectra with peaks at the LL positions until the critical electric field value $E=v_F B$ is reached.

\section*{The case of strain.}

As it is known, in the low energy effective model of graphene, lattice deformations couple to the Dirac fermion as elastic U(1) vector fields that depend linearly on the strain tensor $u_{ij}= (\partial_i u_j +\partial_j u_i + \partial_i h \partial_j h)/2$ as \cite{KM97,SA02b} 
\begin{eqnarray}
A_x &=&\frac{\beta }{2a}\phi_0\left( u_{xx}-u_{yy}\right) ,  \nonumber
\\
A_y &=&-\frac{\beta }a \phi_0 u_{xy}, \label{strainfield}
\end{eqnarray}
where $u_x$, $u_{y}$ and $h$ are in-plane and out-of-plane deformations, respectively, 
 $\beta$ is a dimensionless parameter related to the elastic properties of the material and estimated to be of order 2 in graphene, $a$ is the lattice spacing, and $\phi_0$ is the flux quantum.
The simplest strain configuration giving rise to a constant pseudomagnetic field $B$ provides the elastic vector potential ${\bf A}= (-By, 0)$, which can be reached with $u_{xx}=-2a/(\beta \phi_0)By$ and $u_{yy}=0$. This deformation will break the spectrum into PLL. As discussed in \cite{MJSV13}, strain modifies the effective low energy Hamiltonian in many other ways. The three more relevant perturbations are the just presented vector field coupling, the space--dependent Fermi velocity that we will ignore since it is a higher order effect, and the deformation potential. This term, proportional to the trace of the strain tensor, will be the focus of this work. The deformation potential has been studied at length in graphene \cite{SPG07,HdSacph07,MvO10,COK+10,OCKG11} as  the major cause of electronic scattering from lattice deformations \cite{CXI+07,DYetal10,efetovHighDens}. 

For the simple configuration chosen, the deformation potential  is 
$A_0 = u_{xx}+u_{yy}=-2a/(\beta \phi_0)By $. The effective Hamiltonian to this order in strain is
\beq
H=v_F {\vec \sigma}\cdot({\vec k}+{\vec A})+\lambda A_0\mathbb{I},
\label{ham}
\eeq 
where the coupling constant $\lambda$ will be discussed later. As we see, the deformation potential acts as an electric field $E_y=-\lambda\partial_y A_0=2a \lambda/(\beta \phi_0 ) B$.
We then have formally the same situation as in \cite{LSB07,PC07}. In this particular toy model example, the PLL collapse happens for a critical electric field which is independent on the applied strain, so that the condition $E_c=v_F B$ yields a lower bound on the value of the coupling constant $\lambda$ in \eqref{ham}. For the critical field to be reached from elasticity we then obtain $\lambda \gtrsim v_F \phi_0 \beta/(2a) \sim 4-7 ~\mathrm{eV}$. Experimental values  extracted from the temperature dependence of the resistivity have been shown in \cite{OCKG11} to be compatible with $\lambda \sim 4~ \mathrm{eV}$ \cite{CXI+07,DYetal10,efetovHighDens}, already including screening effects. This shows that the effect is by no means negligible, and that more realistic strain configurations should be considered.   

\begin{figure}
\includegraphics[width=12cm]{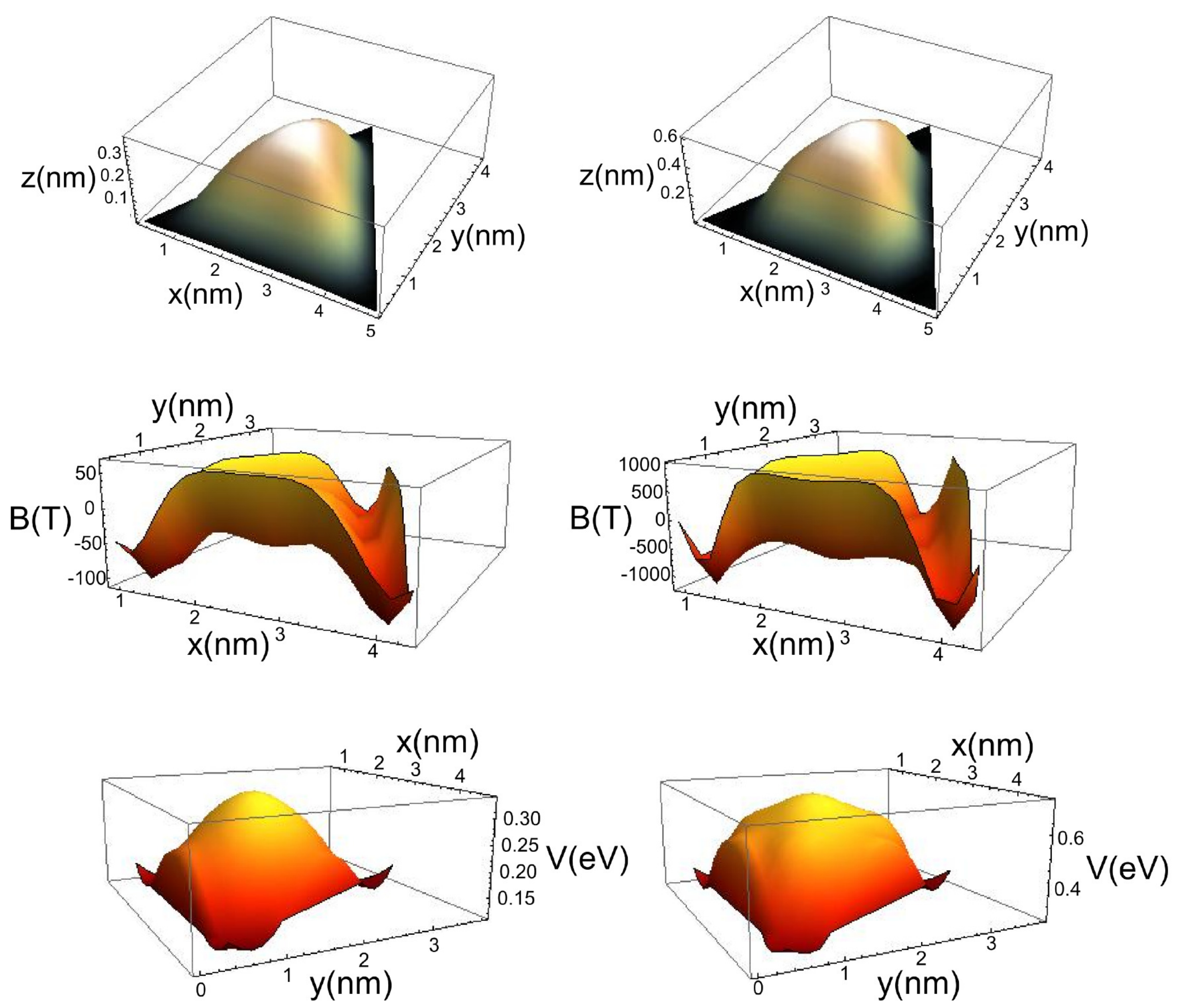}
\caption{Shape (upper), effective magnetic field (middle), and deformation potential (lower) of the two triangular bubbles discussed in \cite{LBetal10}. Figs. taken from ref. \cite{LBetal10} (Supporting material).}
\label{fig:C1}
\end{figure}  
To see the consequences of the effect described in this work, we will re-examine the strain configurations shown in \cite{LBetal10}.  Fig.~\ref{fig:C1} taken from this reference, shows the shape (upper part), effective magnetic field (middle), and deformation potential (lower part) of the two triangular bubbles discussed in \cite{LBetal10}. As we see, the maximum electric field is of the same order in both bubbles, $E \sim 0.1~\mathrm{V/nm}$, as estimated from the ratio between the maximum deformation potential and the typical length of the bubbles. Using the critical  electric field $E_c \approx 10^{-3}B\mathrm{[T]~V/nm}$ for graphene, it is obvious, according to our analysis, that PLL should not be observed in the left figure despite the high pseudomagnetic field of $50~\mathrm{T}$. The shape in the right hand side will show PLL. The difference between the two shapes is that the left one has lower strain of order $4\%$, thus lower pseudomagnetic field, while the right one has higher strain of order $10\%$, which originates much higher pseudomagnetic fields. Our prediction is that it is easier to observe PLL associated to very high pseudomagnetic fields than for not so high pseudomagnetic fields.

The analysis in this work also affects the estimation of the pseudomagnetic fields induced by strain. As can be seen in Fig.~\ref{fig:E1}, the separation between the PLL used to evaluate the pseudomagnetic field changes with the value of the deformation potential. This provides a natural explanation for the spread of data when $(\epsilon(n) - \epsilon_D)/\hbar \omega_c$ is plotted at a given $n$ for different spatial points and different samples (Fig.~2C in Ref.~\cite{LBetal10}), where $\epsilon_D$ is the local shift of the Dirac point also due to the deformation potential. The data spread observed is fully compatible with a strain induced electric field that in some samples goes up to the critical electric field $E \sim E_c$.



Note that shear strain as the one discussed in the original reference \cite{GKG10} will not generate a scalar potential so the PLL spectra will not be affected. Also important is the fact that, due to the extreme anisotropy of the graphene lattice, slightly rotated strain configurations may give rise to totally different STM images \cite{JMV13}.  
The present analysis can serve to understand better the STM images of strained graphene \cite{YTetal11} and the similar problems of interpretation arising at the surface of 3D topological insulators \cite{SCetal16}. Moreover, the change on PLL spectra can equally be induced with an external electric field, instead of that due to the scalar potential. The critical field $E_c \approx 10^{-3}B\mathrm{[T]~V/nm}$ for graphene shows that very high pseudomagnetic fields can be manipulated with not so high in-plane electric fields already available in the lab.

The space--dependent Fermi velocity as well as the other terms coupling strain to the electronic density  discussed in \cite{MJSV13}, will not affect the position of the PLL but  will also change the shape of the spectral lines. A detailed derivation of the influence of the various terms generated by strain STM on the spectra will be shown in a different work. These details do not change the results of the present analysis.
\\

\acknowledgments
M.A.H.V. thanks F. de Juan and Y. Ferreiros for useful conversations.
E.V.C. acknowledges partial support from FCT-Portugal through Grant No.~UID/CTM/04540/2013.
M.A.C. acknowledges the support by the Ministry of Science and
Technology (Taiwan) under contract number NSC 102-
2112-M-007-024-MY5, and Taiwans National Center of
Theoretical Sciences (NCTS).
M.A.H.V.  has been supported by Spanish MECD
grants FIS2014-57432-P and by the European Union Seventh Framework
Programme under grant agreement no. 604391 Graphene Flagship.

\bibliography{SHbib}

\end{document}